\documentclass[a4,article,nofootinbib]{revtex4}

\usepackage{graphicx}
\usepackage{dcolumn}
\usepackage{bm}
\usepackage{color}
\usepackage{amsmath}%
\usepackage{amsfonts}%
\usepackage{amssymb}%
\usepackage{breqn}
\usepackage{fixfoot}
\usepackage{mathrsfs}

\newcommand{\mc}{\mathcal}
\newcommand{\cp}{\times}

\newcommand{\cu}{\nabla\times}

\newcommand{\bol}{\boldsymbol}

\newcommand{\abs}[1]{\left\lvert{#1}\right\rvert}

\newcommand{\lr}[1]{\left({#1}\right)}
\newcommand{\ea}[1]{\left\langle{#1}\right\rangle}

\newcommand{\p}{\partial}

\makeatletter
\let\cat@comma@active\@empty
\makeatother

\begin{document}
\title{Charged Particle Diffusion in a Magnetic Dipole Trap}
\author{N. Sato and Z. Yoshida}
\affiliation{Graduate School of Frontier Sciences, The University of Tokyo,
Kashiwa, Chiba 277-8561, Japan}
\date{\today}

\begin{abstract}
When particles are magnetized, a diffusion process is influenced by the ambient magnetic field. 
While the entropy increases, the constancy of the magnetic moment puts a constraint.
Here, we compare the E-cross-B diffusion caused
by random fluctuations of the electric field in two different systems, the Penning-Malmberg trap and the magnetic dipole trap.
A Fokker-Planck equation is derived by applying the ergodic ansatz on the invariant measure of the system.
In the dipole magnetic field particles diffuse inward and accumulate in the higher magnetic field region, while, in a homogeneous magnetic field, particles diffuse out from the confinement region. 
The properties of analogous transport in a more general class of magnetic fields are also briefly discussed.     
\end{abstract}

\keywords{\normalsize }

\maketitle


\section{INTRODUCTION}

Dipole magnetic fields, which are commonly found in the universe, work as natural confinement devices for charged particles (cf. \cite{bib01}). The underlying mechanism of spontaneous confinement is the adiabatic invariance of the magnetic moment, which puts a topological constraint on the diffusion process of particles across magnetic field lines. 
The same mechanism can be applied to design an effective charged-particle trap \cite{bib02,bib03}.  
The sharp density gradient observed in laboratory experiments \cite{bib04,bib05}
is thought to be generated by a peculiar random walk, the so-called inward diffusion \cite{bib06,bib07}, which tends to homogenize the number of particles contained in each flux-tube \cite{bib01,bib08}. 
The creation of the density gradient is fully consistent with the second
law of thermodynamics because the thermodynamically consistent entropy measure 
is defined on a magnetic coordinate system reflecting the conservation of the first adiabatic invariant throughout the diffusion process \cite{bib09,bib10,bib11}. Then, the confined plasma achieves an equilibrium (maximum entropy) state characterized by a rigid rotation around the symmetry axis \cite{bib02,bib12}, as it happens in the case of a Penning-Malmberg trap \cite{bib13,bib14,bib15,bib16}.
Due to such peculiar properties, 
magnetic dipole traps for antimatter and pair plasma confinement are currently
under development \cite{bib17,bib18} and represent a promising source of applications \cite{bib19}.
 
The aim of the present paper is twofold. 
The first objective is the formulation of 
a thermodynamically consistent diffusion operator in an inhomogeneous magnetic field.
This is achieved by exploiting Liouville's theorem \cite{bib20} of canonical Hamiltonian mechanics and by careful application of the ergodic hypothesis \cite{bib21} on the corresponding invariant measure (the phase space volume element preserved by the dynamical flow), where an $H$-theorem is proved. The ergodic hypothesis on the invariant measure enables perturbations to be represented
in terms of random processes, which determine a Fokker-Planck equation for the plasma density.
The obtained diffusion operator strongly departs from conventional and phenomenological diffusion operators
and reflects the inhomogeneity and the curvature of the magnetic field, as well as the intrinsic properties of the fluctuations.
The second goal of this study is to show that the
transport caused by $\bol{E}\cp\bol{B}$ drift in a dipole magnetic field pushes particles toward regions
of high magnetic field strength, resulting in spontaneous confinement of the plasma.
It is concluded that such kind of magnetic trap should provide a fundamental advantage
when electrostatic fluctuations are taken into account.

\section{ExB DRIFT FROM THE VIEWPOINT OF HAMILTONIAN MECHANICS}

The equation of motion of a charged particle in an electromagnetic field is given by:
\begin{equation}
m\frac{d\bol{v}}{dt}=q\lr{\bol{E}+\bol{v}\cp\bol{B}}.\label{eq1}
\end{equation}
In this notation, $m$, $q$, $\bol{v}=\dot{\bol{x}}$, $t$, $\bol{E}=-\nabla\phi-\partial_{t}\bol{A}$, $\bol{B}=\cu{\bol{A}}$ are particle mass, particle charge, particle velocity, time, electric field and magnetic field respectively. Electric and magnetic fields are represented through the standard scalar and vector potentials $\lr{\phi,\bol{A}}$. 
Equation (\ref{eq1}) can be cast in canonical Hamiltonian form in terms of the Hamiltonian function (representing particle energy):
\begin{equation}
H=\frac{1}{2m}\lr{\bol{p}-q\bol{A}}^{2}+q\phi,\label{eq2}
\end{equation}
where the canonical variables are $\lr{\bol{p},\bol{q}}=\lr{m\bol{v}+q\bol{A},\bol{x}}$.

Purpose of the present study is to investigate the effect of $\bol{E}\cp\bol{B}$ drift
driven diffusion in different magnetic topologies on transport and confinement. 
In the following, the magnetic field will be considered static, 
i.e. $\p_{t}\bol{A}=\bol{0}$ and $\bol{E}=-\nabla\phi$. 
Furthermore, we shall not be concerned with the dynamics occurring at the time and length scales of the fast cyclotron gyration around the magnetic field. 
In order to obtain the $\bol{E}\cp\bol{B}$ drift
velocity, equation (\ref{eq1}) must be `reduced' in the following way.
First, the non-inertial limit $m\rightarrow 0$ is taken.
Physically, this is because when the motion across field lines is considered the mass of a charged particle
is effectively small, while inertial effects mainly contribute to the cyclotron motion and to the dynamics along the magnetic field. The result of the reduction is:
\begin{equation}
\bol{0}=\bol{E}+\bol{v}\cp\bol{B}.\label{eq3}
\end{equation}  
Similarly, the Hamiltonian function becomes non-inertial:
\begin{equation}
H=q\phi.\label{eq4}
\end{equation}
It is worth to mention that equation (\ref{eq3}) is still Hamiltonian \cite{bib22}.

Now observe that in the reduced equation of motion (\ref{eq3}) the electric field is always perpendicular to the magnetic field. Therefore, the charged particle is not subjected to any force along $\bol{B}$. 
This allows a further reduction $\bol{v}_{\parallel}\rightarrow\bol{0}$, where the velocity parallel
to the magnetic field is completely discarded. Then, equation (\ref{eq3}) can be inverted to give
the $\bol{E}\cp\bol{B}$ drift equation of motion:
\begin{equation}
\bol{v}=\frac{\bol{E}\cp\bol{B}}{B^{2}}=\frac{\bol{B}}{B^{2}}\cp\nabla\phi.\label{eq5}
\end{equation}
It can be easily verified that the energy associated to equation (\ref{eq5}) is $q\phi$. Indeed:
\begin{equation}
\dot{\phi}=\nabla\phi\cdot\lr{\frac{\bol{B}}{B^{2}}\cp\nabla\phi}=0.\label{eq6}
\end{equation}
However, the $\bol{E}\cp\bol{B}$ drift equation of motion (\ref{eq5}) is not, in general, Hamiltonian.
As it will be discussed in the following section, the construction of a statistical theory of diffusion
requires the existence of an invariant measure (a volume element preserved by the dynamical flow).
In the standard formulation of statistical mechanics such volume element is provided by Liouville's theorem of canonical Hamiltonian mechanics. Therefore, it is crucial to determine the condition under which
equation (\ref{eq5}) admits an Hamiltonian form. Such condition is given by the so-called Jacobi identity
\cite{bib23}, which must be satisfied by the Poisson operator of the system. In the present case 
the candidate Poisson operator is represented by the vector field $\bol{B}/B^{2}$, and the Jacobi identity reads:
\begin{equation}
\frac{\bol{B}}{B^{2}}\cdot\nabla\cp\lr{\frac{\bol{B}}{B^{2}}}=\frac{\bol{B}\cdot\nabla\cp\bol{B}}{B^{4}}=0.\label{eq7}
\end{equation}
Clearly, equation (\ref{eq7}) is not satisfied by a general magnetic field.
In three dimensions, the condition (\ref{eq7}) has a geometrical interpretation:
a smooth non-vanishing vector field obeying (\ref{eq7}) is said to be integrable (in the sense of the Frobenius theorem \cite{bib24}) and, as such, admits the local representation $\bol{B}=\lambda\nabla C$ for some scalar functions $\lambda$ and $C$. 
Hence, the $\bol{E}\cp\bol{B}$ drift equation of motion (\ref{eq5}) is Hamiltonian
as long as the magnetic field is integrable.

\begin{figure}[h]
\hspace*{-1.4 cm}
\centering
\includegraphics[scale=0.18]{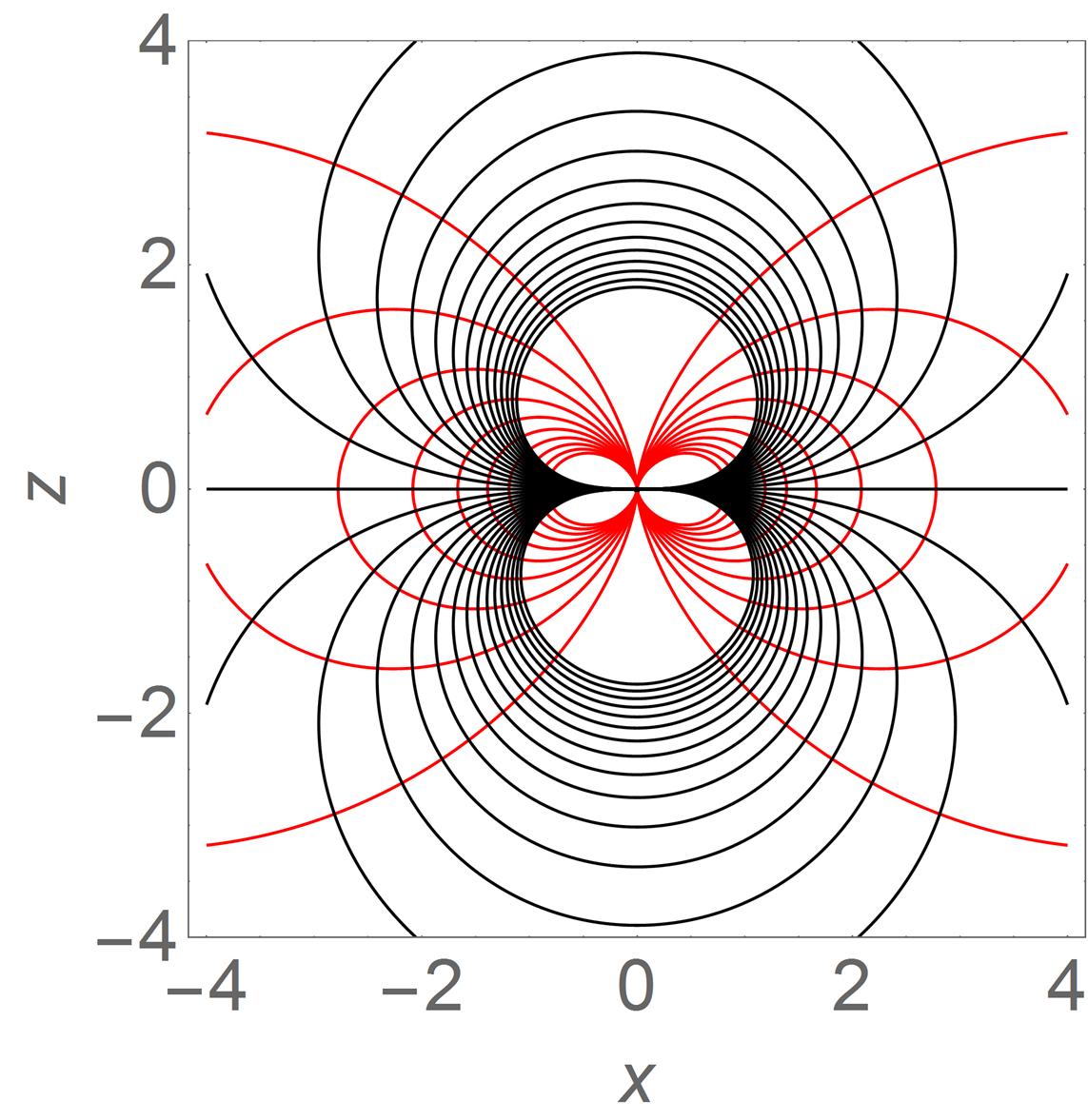}
\caption{\footnotesize Contour plot of $\psi$ (red contours) and $C$ (black contours) for the point dipole magnetic field (\ref{eq9}) in the plane $y=0$.}
\label{fig1}
\end{figure}

Both the straight magnetic field of a Penning-Malmberg trap and
the dipole magnetic field of a magnetic dipole trap satisfy the vacuum condition $\cu\bol{B}=\bol{0}$ in those spatial regions that are accessible to the plasma (in the dipole case $\cu\bol{B}\neq\bol{0}$ only within the coil generating the magnetic field). Therefore, in the domain of interest, equation (\ref{eq7}) is satisfied and the corresponding $\bol{E}\cp\bol{B}$ drift is Hamiltonian. Furthermore, notice that
in both cases $\bol{B}=\nabla C$, implying that $C$ is a constant of motion (a so-called Casimir invariant):
\begin{equation}
\dot{C}=\nabla C\cdot\lr{\frac{\nabla C}{\abs{\nabla C}^{2}}\cp\nabla\phi}=0.\label{eq8}
\end{equation}
It is useful to give $C$ explicitly. In a straight magnetic field aligned
to the $z$-axis of a cylindrical coordinate system $\lr{r,z,\theta}$, the Casimir invariant is $C=z$. In the case of a point dipole magnetic field \cite{bib09}, 
\begin{equation}
\bol{B}=\nabla\psi\cp\nabla\theta,\label{eq9}
\end{equation}
where $\psi=r^{2}\lr{r^{2}+z^{2}}^{-3/2}$ is the flux function, one finds that $C=z\lr{r^{2}+z^{2}}^{-3/2}$.
Notice that physical units were omitted. A contour plot of $\psi$ and $C$ is given in figure \ref{fig1}.

While the discussion of non-Hamiltonian diffusion in non-integrable magnetic fields is beyond the scope of the present paper,
some comments on the resulting statistical properties will be given in the conclusion.

\section{A STATISTICAL THEORY OF DIFFUSION}
\subsection{Invariant Measure}
In the previous section it was shown that in straight and dipole magnetic fields $\bol{E}\cp\bol{B}$ dynamics
(\ref{eq5}) is Hamiltonian. Therefore, in virtue of Liouville's theorem, an invariant measure $dV_{I}$ can be found. Let $J$ be the Jacobian of the coordinate change from the Cartesian reference system $\bol{x}=\lr{x,y,z}$ 
to the coordinates spanning $dV_{I}$: 
\begin{equation}
dV_{I} = J\,dV = J\,dx\,dy\,dz.\label{eq10}
\end{equation}  
The next step required to construct a diffusion operator is the
explicit identification of $J$, which is in general a function of the spatial variables $\bol{x}$. 
By definition, the volume element $dV_{I}$ is said to be invariant with respect to
the dynamical flow $\bol{v}$ whenever 
\begin{equation}
\nabla\cdot\lr{J\bol{v}}=0.\label{eq11}
\end{equation}  
Substituting equation (\ref{eq5}), and recalling that in the case under examination $\bol{B}=\nabla C$, we arrive at the condition:
\begin{equation}
\nabla\phi\cdot\left[\nabla\lr{\frac{J}{B^{2}}}\cp\nabla C\right]=0.\label{eq12}
\end{equation}
The solution of the nontrivial case $C\neq\phi$ is:
\begin{equation}
J=B^{2} \mc{F}\lr{\phi,C},\label{eq13}
\end{equation}
where $\mc{F}$ is an arbitrary scalar function of $\phi$ and $C$.
However, for the statistical theory to be independent of the specific properties of matter, 
the invariant measure cannot depend on the Hamiltonian $q\phi$. Therefore, $\mc{F}$ is chosen 
to be only a function of the Casimir invariant $C$, i.e. $\mc{F}=\mc{F}\lr{C}$.
Notice that, in light of equation (\ref{eq13}), the magnetic energy contained
in a volume $\Omega$ transported with velocity given by equation (\ref{eq5}) is always preserved,
$\dot{E}_{M}=\frac{d}{dt}\int_{\Omega}\frac{B^{2}}{2\mu_{0}}\,dV=0$ where $\mu_{0}$ is the vacuum permeability.

\subsection{Ergodic Hypothesis}

Using the invariant measure $dV_{I}$, the ergodic hypothesis can now be applied to
obtain a convenient expression of the turbulent electric field $\delta\bol{E}$ in terms of random processes. First, define the perturbation $\delta\bol{E}$ as the
departure of the total electric field $\mc{E}$ from the `macroscopic' term $\bol{E}=-\nabla\phi$ as a result of random interactions among charged particles in the ensemble:
\begin{equation}
\mc{E}=\bol{E}+\delta\bol{E}=-\nabla\phi+\delta\bol{E}.\label{eq14}
\end{equation}
By definition,
\begin{equation}
\ea{\delta\bol{E}}=\ea{\mc{E}-\bol{E}}=\bol{0}.\label{eq15}
\end{equation}
Notice that the ensemble average $\ea{\,\cdot\,}$ is carried out on the invariant measure $dV_{I}$:
\begin{equation}
\ea{\bol{E}}=\int_{\Omega}{f\bol{E}}\,dV_{I},\label{eq16}
\end{equation}
where $f$ is the normalized distribution function 
such that $dN=fdV_{I}$ corresponds to the particle number contained in the volume element, 
and $\Omega$ is the volume occupied by the plasma with $\int_{\Omega}{f}\,dV_{I}=1$.

In virtue of the fact that $dV_{I}$ is an invariant measure, whose existence
is a necessary condition of the ergodic theorem \cite{bib21}, the ergodic hypothesis can be enforced and ensemble averages can be exchanged with time averages:
\begin{equation}
\ea{\delta\bol{E}}=\lim_{T\rightarrow\infty}\frac{1}{T}\int_{0}^{T}{\delta\bol{E}}\,dt=\bol{0}.\label{eq17}
\end{equation} 
Here, $T$ is a time interval and equation (\ref{eq15}) was used. 
Hence, the time average of the perturbation $\delta\bol{E}$ over a sufficiently large $T$ must vanish. It is only under these conditions that $\delta\bol{E}$ can be suitably represented
by a random process of zero time average:
\begin{equation}
\delta\bol{E}=-D^{1/2}\bol{\Gamma},\label{eq18}
\end{equation} 
where $\bol{\Gamma}=\lr{\Gamma_{x},\Gamma_{y},\Gamma_{z}}$ is a three dimensional Gaussian white noise process and the constant and positive parameter $D^{1/2}$, bearing physical units of $N\,C^{-1}\,s^{1/2}$, scales the strength of fluctuations.
It is important to stress again that, without the invariant measure, there would not
be any statistical justification of the representation (\ref{eq18}). Furthermore,
observe that the distribution function $f$ characterizing the ensemble is defined on
$dV_{I}$, and not on the Cartesian reference system $\lr{x,y,z}$. It will be shown soon that
this second fact has fundamental implications for the consistency of the diffusion process
with the second law of thermodynamics.

\subsection{Fokker-Planck Equation}
This paragraph is dedicated at the derivation of the Fokker-Planck equation satisfied
by the distribution function $f$. 
First, notice that the total effective electric field $\tilde{\mc{E}}$ acting on a charged particle is now:
\begin{equation}
\tilde{\mc{E}}=-\nabla\phi-D^{1/2}\bol{\Gamma}-\frac{\gamma}{2q}\bol{v}.\label{eq19}
\end{equation}
The last term in the equation is a friction force that is proportional (through the friction coefficient $\gamma>0$) to the velocity (\ref{eq5}) as in the conventional definition. Such friction term, which counterbalances
the self-induced fluctuations $\delta\bol{E}$, must be included to ensure preservation of
total energy $\mc{H}=\int_{\Omega}fq\phi\,dV_{I}$. 
Thus, the stochastic equation of motion satisfied by a charged particle in the
turbulent electric field (\ref{eq19}) is:
\begin{equation}
\mc{\bol{V}}=\dot{\bol{x}}=\frac{\bol{B}}{B^{2}}\cp\lr{\nabla\phi+D^{1/2}\bol{\Gamma}+\frac{\gamma}{2q}\bol{v}}.\label{eq20}
\end{equation}
Following the standard procedure (see \cite{bib06,bib07,bib25}), equation (\ref{eq20})
can be converted to a Fokker-Planck equation for the particle density $\rho$ 
in the Cartesian reference frame $\lr{x,y,z}$ where the stochastic velocity $\mc{\bol{V}}$ is defined. The Fokker-Planck equation reads \cite{bib26}:
\begin{equation}
\frac{\p \rho}{\p t}=\nabla\cdot\left[-\rho\lr{\bol{v}+\frac{\gamma\bol{B}}{2qB^{2}}\cp\bol{v}}
+\frac{1}{2}D\,\frac{\bol{B}}{B^{2}}\cp\cu\lr{\frac{\bol{B}}{B^{2}}\rho}\right].\label{eq21}
\end{equation}
Observe how the diffusion operator of $\bol{E}\cp\bol{B}$ drift dynamics 
(the term proportional to $D$ in the equation) strongly depends on the topology of the magnetic field. 

Finally, it must be emphasized that while the present derivation was limited to
pure $\bol{E}\cp\bol{B}$ drift to capture its role in plasma transport, 
the general Fokker-Planck equation accounting for the full dynamical picture can be obtained
by application of the same procedure (see \cite{bib06,bib07}).


\subsection{H-Theorem}
Purpose of the present section is to show that the Fokker-Planck equation (\ref{eq21})
satisfies Boltzmann's $H$-theorem on the invariant measure $dV_{I}$, and it is therefore consistent with the second law of thermodynamics. As a rewarding result, the form of the distribution function at equilibrium will be obtained.

Let $\rho$ be the (normalized) particle density in the Cartesian reference frame $\lr{x,y,z}$.
Evidently:
\begin{equation}
fdV_{I}=\rho dV.\label{eq22}
\end{equation}
Hence, 
\begin{equation}
\rho=fJ=fB^{2}\mc{F}\lr{C}.\label{eq23}
\end{equation}
On the other hand, the entropy measure $\Sigma$ of the system must be defined
on the invariant measure $dV_{I}$ according to:
\begin{equation}
\Sigma=-\int_{\Omega}f\log{f}\,dV_{I}=-\int_{\Omega}{\rho\log\lr{\frac{\rho}{J}}}\,dV.\label{eq24}
\end{equation}
Here, equations (\ref{eq22}) and (\ref{eq23}) were used.
Observe that the entropy measure $\Sigma$ departs from the 
conventional measure $S=-\int_{\Omega}{\rho\log{\rho}}\,dV$ due to the Jacobian $J$.
That $\Sigma$ is the correct entropy measure can be verified by evaluating its rate of change.
For this purpose, take the domain $\Omega$ to be such that 
$\bol{B}\cp\lr{\nabla f\times\bol{B}}\cdot\bol{n}=\bol{0}$ on the boundary $\p\Omega$, where $\bol{n}$ is the outward normal to $\p\Omega$, and assume the boundary condition $\nabla\phi=\bol{0}$ on $\p\Omega$ (these conditions ensure that there is no probability flowing out of the prescribed domain so that the system is thermodynamically isolated). 
Recalling that $\bol{B}=\nabla C$, $\mc{F}=\mc{F}\lr{C}$, and using boundary conditions and equation (\ref{eq21}):  
\begin{dmath}
\frac{d\Sigma}{dt}=-\int_{\Omega}{\p_{t}\rho\lr{1+\log{\lr{\frac{\rho}{J}}}}}\,dV
=\int_{\Omega}{\left[-\rho\lr{\bol{v}+\frac{\gamma\bol{B}}{2qB^{2}}\cp\bol{v}}
+\frac{1}{2}D\,\frac{\bol{B}}{B^{2}}\cp\cu\lr{\frac{\bol{B}}{B^{2}}\rho}\right]\cdot\nabla\lr{\log{\lr{\frac{\rho}{J}}}}}\,dV
=\frac{D}{2}\int_{\Omega}{\rho
\left[\frac{\gamma}{qD}\nabla\phi+\nabla{\log{\lr{\frac{\rho}{B^{2}}}}}\right]\cp\frac{\bol{B}}{B^{2}}
\cdot\left[\nabla\log\lr{\frac{\rho}{B^{2}}}\cp\frac{\bol{B}}{B^{2}}\right]}\,dV
=\frac{D}{2}\int_{\Omega}{\rho\abs{\nabla\lr{\frac{\gamma\phi}{qD}+\log\lr{\frac{\rho}{B^{2}}}}\cp\frac{\bol{B}}{B^{2}}}^{2}}\,dV
-\frac{\gamma}{2q}\int_{\Omega}{\rho\left[\nabla\lr{\frac{\gamma\phi}{qD}+\log\lr{\frac{\rho}{B^{2}}}}\cp\frac{\bol{B}}{B^{2}}\right]\cdot\left[\nabla\phi\cp\frac{\bol{B}}{B^{2}}\right]}\,dV.\label{eq25}
\end{dmath}
 On the other hand, conservation of total energy demands that:
\begin{dmath}
\frac{d\mc{H}}{dt}=\int_{\Omega}{\p_{t}\rho q\phi}\,dV=
q\int_{\Omega}{\left[\rho\lr{\bol{v}+\frac{\gamma\bol{B}}{2qB^{2}}\cp\bol{v}}
-\frac{1}{2}D\,\frac{\bol{B}}{B^{2}}\cp\cu\lr{\frac{\bol{B}}{B^{2}}\rho}\right]\cdot\nabla\phi}\,dV
=-\frac{qD}{2}\int_{\Omega}{\rho\left[\nabla\lr{\frac{\gamma\phi}{qD}+\log{\lr{\frac{\rho}{B^{2}}}}}\cp\frac{\bol{B}}{B^{2}}\right]\cdot\left[\nabla\phi\cp\frac{\bol{B}}{B^{2}}\right]}\,dV
=0.
\end{dmath}
Substituting this result in equation (\ref{eq25}), it follows that:
\begin{equation}
\frac{d\Sigma}{dt}=\frac{D}{2}\int_{\Omega}{\rho\abs{\nabla\lr{\frac{\gamma\phi}{qD}+\log\lr{\frac{\rho}{B^{2}}}}\cp\frac{\bol{B}}{B^{2}}}^{2}}\,dV\geq0.\label{eq27}
\end{equation}
Therefore, Boltzmann's H-theorem is satisfied.
Furthermore, in the limit $t\rightarrow\infty$ the rate of change in $\Sigma$ must vanish: 
\begin{equation}
\lim_{t\rightarrow\infty}\frac{d\Sigma}{dt}=0.\label{eq28}
\end{equation} 
Then, assuming $\rho>0$ and noting that the integrand in equation (\ref{eq27}) is strictly positive, equation (\ref{eq28}) implies:
\begin{equation}
\lim_{t\rightarrow\infty}\left[\nabla\lr{\frac{\gamma\phi}{qD}+\log\lr{\frac{\rho}{B^{2}}}}\cp\nabla C\right]=\bol{0}\,\,\,\,a.e.\label{eq29}
\end{equation}
Now define $\rho_{\infty}=\lim_{t\rightarrow\infty}\rho$ and further assume that $\rho_{\infty}\in C^{2}\lr{\Omega}$. 
From equation (\ref{eq29}) one obtains:
\begin{equation}
\rho_{\infty}=Z_{\infty}^{-1}B^{2}\exp\left\{-\beta \left[q\phi+ \alpha G\lr{C}\right]\right\},\label{eq30}
\end{equation} 
where $Z_{\infty}^{-1}>0$ is a constant with units of $m^{-3}T^{-2}$, $\alpha$ a constant with units of $s^{-1}$, $G\lr{C}$ an arbitrary function of the Casimir invariant $C$ with units of $J s$, and $\beta=\gamma q^{-2} D^{-1}$ the inverse temperature. The function $G$ must be physically determined by requiring that the argument of the exponential, which represents an effective `particle number', satisfies the additive property. Then, equation (\ref{eq30}) can be interpreted as the probability distribution of a grand canonical ensemble, $G$ represents the `action' associated to the invariant $C$, and the `frequency' $\alpha$ the corresponding chemical potential. 

First, consider the implications of result (\ref{eq30}) for a Penning-Malmberg trap.
In such case, $C=z$ and $B=B_{0}$ with $B_{0}$ a positive constant.
Furthermore, in a standard configuration $\p_{z}\rho_{\infty}=0$, giving:
\begin{equation}
\rho_{\infty}=Z_{\infty}^{-1}B^{2}_{0}\exp\left\{-\beta q\phi\right\},\label{eq31}
\end{equation}  
which reduces to the flat density $\rho^{\infty}=Z_{\infty}^{-1}B^{2}_{0}$ for a neutral plasma $\phi=0$.

Next, consider a point dipole magnetic field. Recalling that $C=z\lr{r^{2}+z^{2}}^{-3/2}$, now $B^{2}=B^{2}\lr{r,z}=\lr{r^{2}+4z^{2}}\lr{r^{2}+z^{2}}^{-4}$. Then,
\begin{equation}
\rho_{\infty}=Z_{\infty}^{-1}B^{2}\lr{r,z}\exp\left\{-\beta\left[ q\phi+\alpha G\right]\right\}=Z_{\infty}^{-1}M^{2}\frac{r^{2}+4z^{2}}{\lr{r^{2}+z^{2}}^{4}}\exp\left\{-\beta\left[ q\phi+\alpha G\right]\right\},\label{eq32}
\end{equation}
which is a strongly peaked profile toward the center of the dipole that can confine even a neutral plasma
with density $\rho_{\infty}=c_{0}B^{2}\lr{r,z}\exp\left\{-\beta\alpha G\right\}$. Here, $M$ is a physical constant with units of $T m^{3}$.   
Hence, the result (\ref{eq30}) clearly shows that the topology of the magnetic field 
can generate a steep density gradient as a result of $\bol{E}\cp\bol{B}$ drift.

Finally, one may wonder how equation (\ref{eq30}) changes when the full guiding center dynamics is taken into account.
In such case, the invariant measure is the $6$-dimensional volume element 
$dV_{I}=d\ell d\psi d\theta dv_{\parallel}d\mu d\theta_{c}=m^{-2}B dx dy dz dv_{\parallel}d\mu d\theta_{c}=m^{-2}BdV$.
Here, $\ell$ is the length coordinate along field lines, $v_{\parallel}$ the velocity along the magnetic field, $\mu$ the magnetic
momentum, and $\theta_{c}$ the phase of the cyclotron gyration. 
Now the Jacobian $J$ is the field strength $B$ and the Casimir invariant is $C=\mu$, leading to 
a distribution function $\tilde{f}=fB$ on $dV$ of the type $\tilde{f}=Z_{\infty}^{-1}B\exp\left\{-\beta\left[ H +\alpha \mu\right]\right\}$,
with $Z_{\infty}^{-1}$ a constant in appropriate units and $H=\mu B+\frac{m}{2}v^{2}_{\parallel}+q\phi$ the Hamiltonian function. The reader is referred to \cite{bib06,bib07,bib11} for further details.  

\section{NUMERICAL TEST IN A DIPOLE MAGNETIC FIELD}

\begin{figure}[t]
\centering
\includegraphics[scale=0.2]{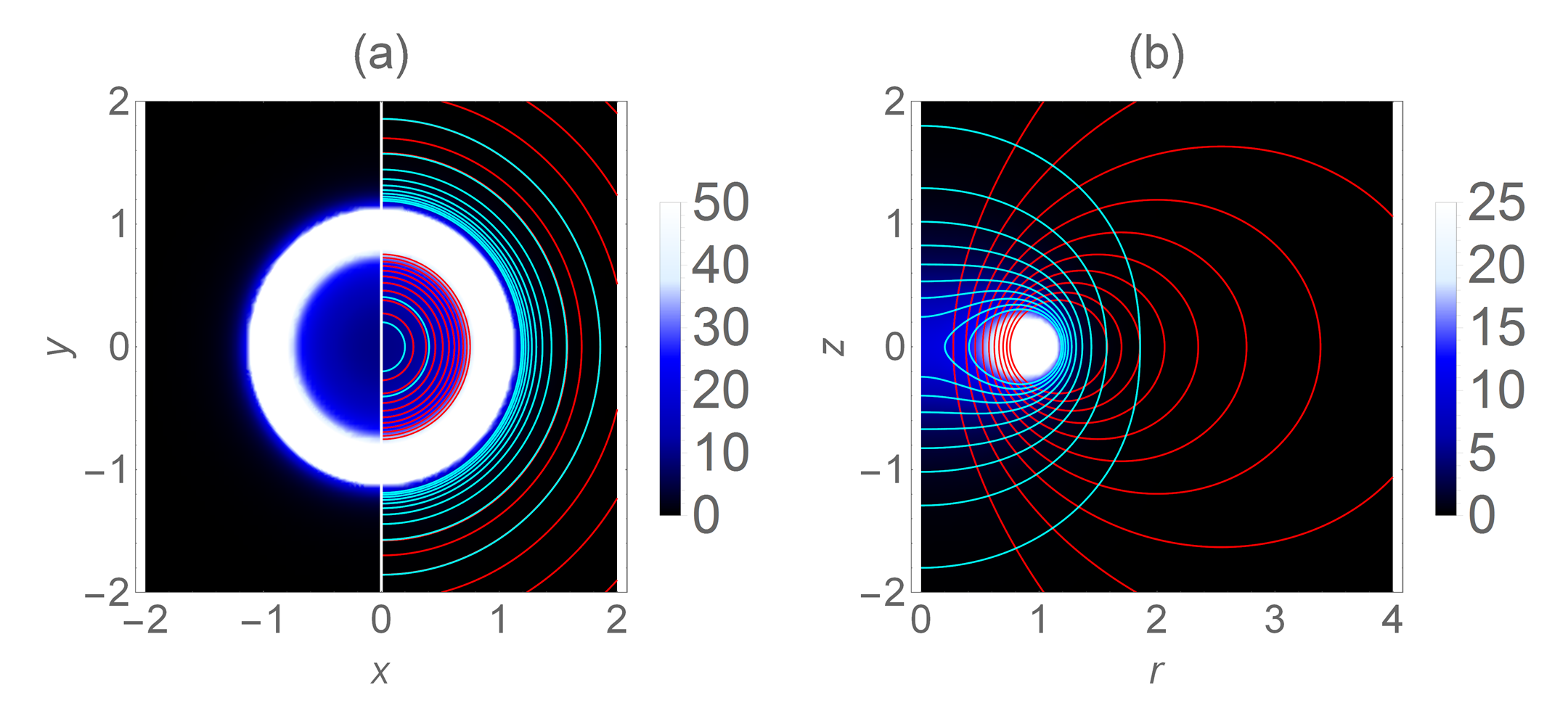}
\caption{\footnotesize Density plot of $B^2$ for the dipole magnetic field (\ref{eq34}). Red lines are the contours of the flux function $\psi$ and represent magnetic field lines. Cyan contours correspond to the level sets of magnetic field strength $B$. (a) Density plot of the function $B^2\lr{x,y,0}$. (b) Density plot of the function $B^{2}\lr{r,z}$.}
\label{fig2}
\end{figure}

\begin{figure}[t]
\hspace*{-1 cm}
\centering
\includegraphics[scale=0.45]{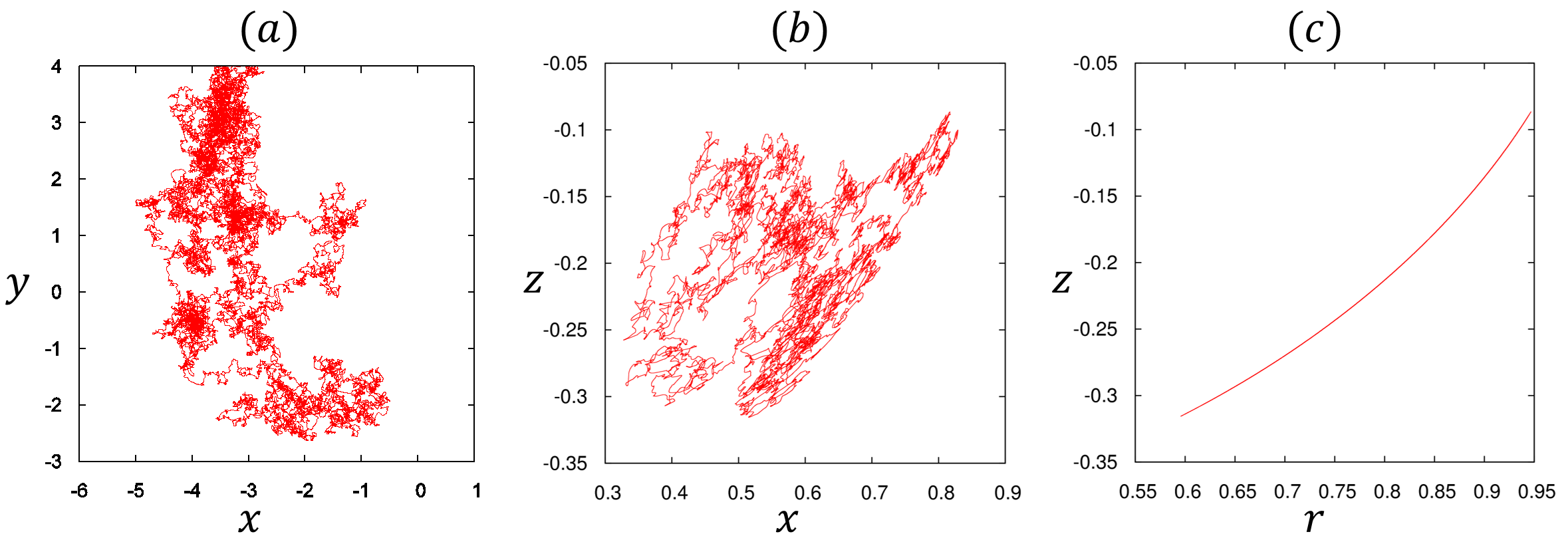}
\caption{\footnotesize (a) Orbit of a particle in the plane $z=0$ for $\bol{E}\cp\bol{B}$ diffusion in a straight magnetic field. (b) and (c): orbit of a particle in the planes $y=0$ and $\lr{r,z}$ for $\bol{E}\cp\bol{B}$ diffusion in a dipole magnetic field. Notice that both orbits lie on the surface $C=$constant, with $\bol{B}=\nabla C$.}
\label{fig3}
\end{figure}

In this section, the theory is put to the test of numerical simulations.
To capture the role of the magnetic field in the diffusion process,
the case of a neutral plasma $\phi=0$ is considered.
The stochastic equation of motion (\ref{eq20}) reduces to:
\begin{equation}
\mc{\bol{V}}=D^{1/2}\frac{\bol{B}}{B^{2}}\cp\bol{\Gamma}.\label{eq33}
\end{equation} 
Equation (\ref{eq33}) will be simulated for an ensemble of $8\cdot 10^{6}$ particles in the
straight magnetic field case, and for an ensemble of $3.4\cdot 10^{5}$ particles in the dipole magnetic field case (the
smaller particle number is due to higher computational cost).
Typical orbits are shown in figure \ref{fig3}.
Then, the resulting density will be compared with equation (\ref{eq30}).
The initial condition is a flat density distribution in a cubic domain centered
at the origin of $\lr{x,y,z}$ space for the straight magnetic field case and a flat density distribution
within the level set $\psi_{0}=0.6$ in the dipole magnetic field case.
If $\bol{B}=\nabla z$ is a straight magnetic field, a flat equilibrium density
profile $\rho_{\infty}=Z_{\infty}^{-1}B_{0}^{2}$ is expected (recall equation (\ref{eq31})).
If $\bol{B}$ is a dipole magnetic field, an inhomogeneous density profile
$\rho_{\infty}=Z_{\infty}^{-1}B^{2}\lr{r,z}\exp\left\{-\beta\alpha G\right\}$ is expected (recall equation (\ref{eq32})). 
In the simulation, instead of the point dipole approximation, 
the magnetic field generated by a current loop
of infinitesimal section and finite radius will be used.
The dipole magnetic field is now written as:
\begin{equation}
\bol{B}=\frac{1}{\sqrt{Q}}\left\{\frac{xz}{r^{2}}\left[\frac{Q-2r}{Q-4r}E\lr{k^{2}}-K\lr{k^{2}}\right],
\frac{yz}{r^{2}}\left[\frac{Q-2r}{Q-4r}E\lr{k^{2}}-K\lr{k^{2}}\right],
\left[\frac{2\left(1+r\right)-Q}{Q-4r}E\lr{k^{2}}+K\lr{k^{2}}\right]\right\}.\label{eq34}
\end{equation}
In this expression, $K$ and $E$ are the complete elliptic integrals of first and second kind,
$k^{2}=4r/Q$, $Q=\lr{1+r}^{2}+z^{2}$.
Furthermore, the magnetic field has been normalized to the 
reference magnetic field $B_{0}=\mu_{0}I/2\pi a$, where $I$ is the current flowing in a loop of radius $a$ (the point dipole
magnetic field can be obtained in the limit $a\rightarrow 0$).
Similarly, the coordinates $\lr{x,y,z}$ are normalized in units of loop radius $a$.
A density plot of $B^{2}$ is given in figure \ref{fig2}.

\begin{figure}[h]
\hspace*{-0.6 cm}
\centering
\includegraphics[scale=0.39]{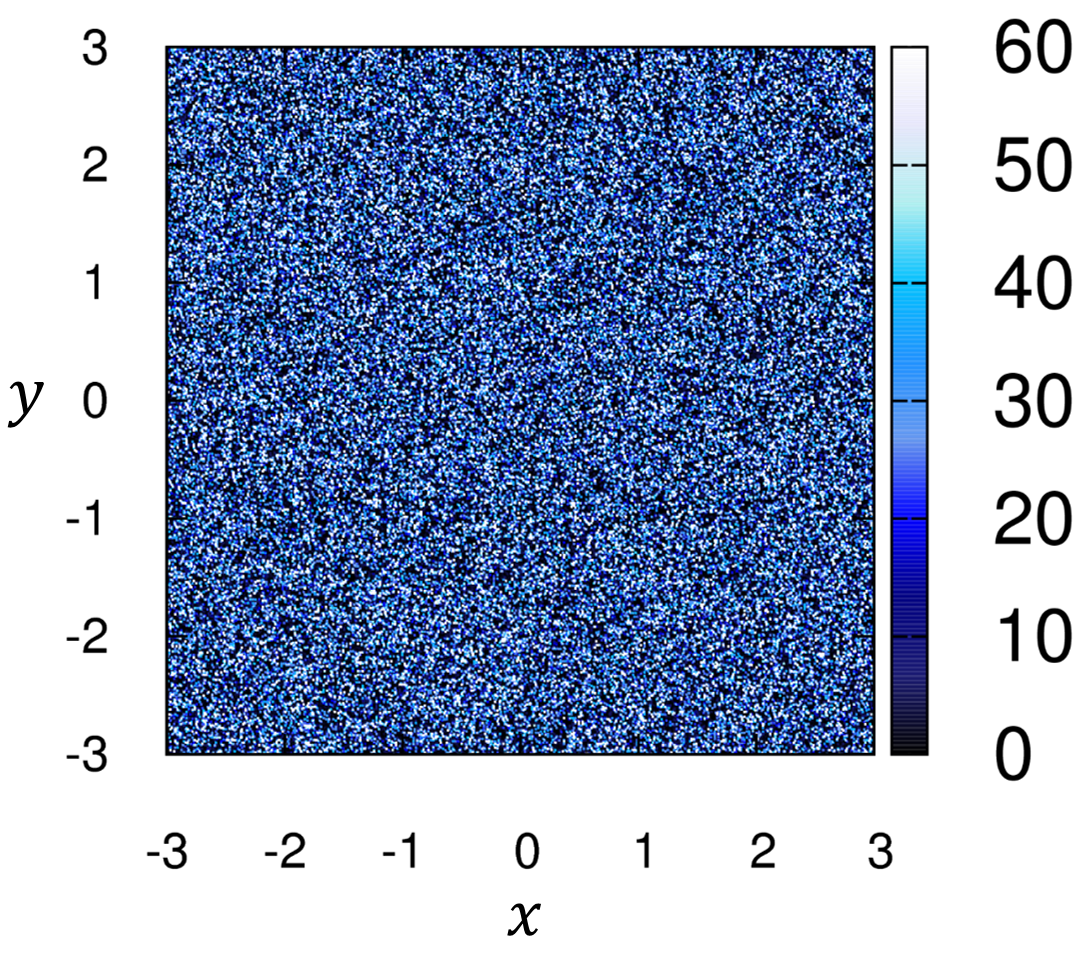}
\caption{\footnotesize Numerically calculated equilibrium density profile in the $\lr{x,y}$ plane for $\bol{E}\cp\bol{B}$ diffusion in a straight magnetic field.}
\label{fig4}
\end{figure}

\begin{figure}[h!]
\centering
\includegraphics[scale=0.35]{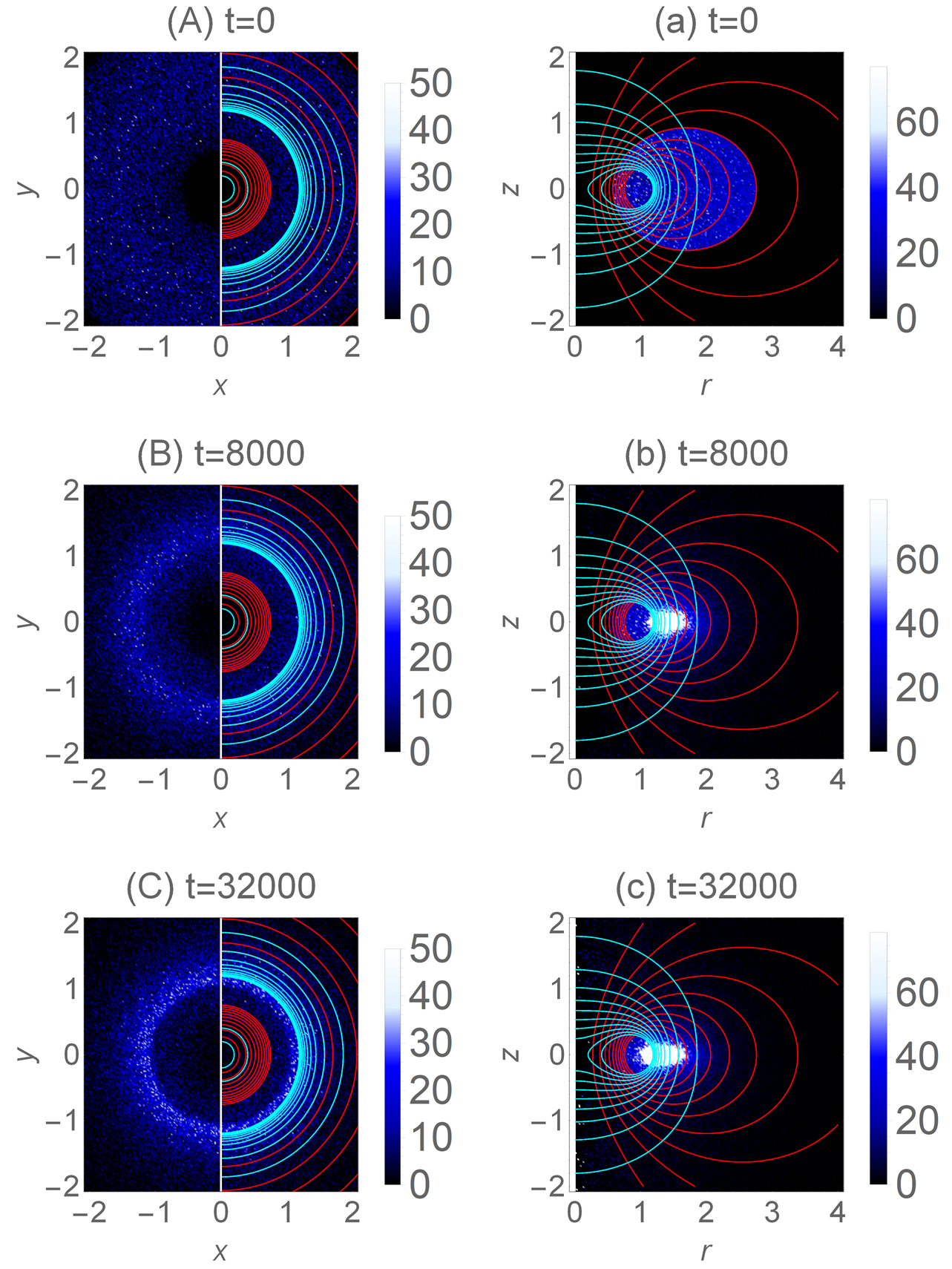}
\caption{\footnotesize Time evolution of density profile for $\bol{E}\cp\bol{B}$ diffusion in a dipole magnetic field.
The left column (plots (A), (B), and (C)) shows the behavior of $u\lr{x,y}=\int{\rho\lr{x,y,z}}\,dz$.
The right column (plots (a), (b), and (c)) shows the behavior of $\int{\rho\lr{r,z}}\,d\theta=2\pi\rho\lr{r,z}$.
Time $t$ is given in arbitrary units. Contours of $\psi$ and $B$ are shown as in figure \ref{fig1}. Additional details are given in the text.}
\label{fig5}
\end{figure}

The result of the numerical simulation in a straight magnetic field is shown in figure \ref{fig4}. As predicted, the distribution is flat.
Figure \ref{fig5} shows the time evolution of the density profile in the case of diffusion
in a dipole magnetic field. In the left column of figure \ref{fig5}, the value $u\lr{x,y}=\int{\rho\lr{x,y,z}}\,dz$ is shown as a function of time $t$. In the right column of figure \ref{fig5}, the
value $\int{\rho\lr{r,z}}\,d\theta=2\pi\rho\lr{r,z}$ is shown as a function of time $t$. 
Notice how $\rho$ progressively approaches $B^2$ of figure \ref{fig2} as expected (the left column of figure \ref{fig4} should be compared with figure \ref{fig2}(a) and the right column with figure \ref{fig2}(b)).
The discrepancy between the profiles of $\rho$ and $B^{2}$ is due to the finiteness of the calculated time interval
and to the chosen initial condition $\rho_{0}=\rho\lr{t=0}=Z_{0}^{-1}\theta\lr{\psi-\psi_{0}}$, with $Z_{0}^{-1}>0$ a constant, 
which is such that $G\lr{C}\neq 0$ in equation (\ref{eq30}). Indeed, since the diffusion process cannot redistribute particles
among different level sets of $C$, the number $dN\lr{C}$ of particles between $C$ and $C+dC$ is always constant, implying
$dN_{0}=2\pi Z_{0}^{-1} dC\int_{\psi_{0}}^{\infty}{\frac{d\psi}{B^2}}=dN_{\infty}=2\pi Z_{\infty}^{-1}e^{-\beta\alpha G} dC\int_{0}^{\infty}{d\psi}$, and therefore $G\lr{C}=-\frac{1}{\beta\alpha}\log{\lr{k\int_{\psi_{0}}^{\infty}{\frac{d\psi}{B^{2}}}}}$ with $k=Z_{\infty}Z_{0}^{-1}/\int_{0}^{\infty}d\psi$. 

\section{CONCLUSION}
In the present paper, the role of $\bol{E}\cp\bol{B}$ drift transport driven by self-induced
electromagnetic fluctuations in integrable magnetic topologies was investigated.
A Fokker-Planck equation was formulated by enforcing the ergodic hypothesis on the invariant measure provided by Liouville's theorem of Hamiltonian mechanics.
Boltzmann's H-theorem was demonstrated, and the equilibrium distribution obtained.
The derived distribution shows that, in the presence of an inhomogeneous magnetic field, 
$\bol{E}\cp\bol{B}$ drift dynamics can confine neutral and non-neutral plasmas by creating a sharp density gradient. It is concluded that inhomogeneous magnetic fields, such as dipole magnetic fields, should be advantageous with respect to plasma confinement in the absence of external driving if compared to homogeneous magnetic configurations.   

The results discussed here pertain to integrable magnetic fields, which make
$\bol{E}\cp\bol{B}$ drift dynamics Hamiltonian.
However, it is of physical interest to elucidate 
how the diffusion process is modified by the introduction of non-integrable magnetic fields,
such as the Beltrami field arising in Taylor's relaxation theory.
Due to the loss of the Hamiltonian structure that enabled the construction of the diffusion operator, the generalization of the theory to arbitrary magnetic fields
requires a fundamental change of perspective that was not addressed here.
Nevertheless, it is worth to mention that beyond integrable magnetic fields
the discriminant between confinement and disruption of the density profile 
is provided by Beltrami fields: in a Beltrami field of uniform strength, the density distribution
will become flat as a result of the diffusion process.

\newpage

\section{ACKNOWLEDGMENTS}
The research of N. S. was supported by JSPS KAKENHI Grant No. 16J01486.



\begin{thebibliography}{99}
\bibitem{bib01} M. Schulz and L. J. Lanzerotti, \textit{Particle Diffusion in the Radiation Belts} (Springer, New York, 1974).
\bibitem{bib02} Z. Yoshida, H. Saitoh, J. Morikawa, S. Watanabe, and Y. Ogawa, Phys. Rev. Lett. \textbf{104}, 235004 (2010).
\bibitem{bib03} H. Saitoh, Z. Yoshida, C. Nakashima, H. Himura, J. Morikawa, and M. Fukao, Phys. Rev. Lett. \textbf{92}, 25 (2004).
\bibitem{bib04} Z. Yoshida, H. Saitoh, Y. Yano, H. Mikami, N. Kasaoka, W. Sakamoto, J. Morikawa, M. Furukawa, and S. M. Mahajan, Plasma Phys. Control. Fusion \textbf{55}, 014018 (2013).
\bibitem{bib05} A. C. Boxer, R. Bergmann, J. L. Ellsworth, D. T. Garnier, J. Kesner, M. E. Mauel, and P. Woskov, Nature Phys. \textbf{6}, pp. 207-212 (2010).
\bibitem{bib06} N. Sato and Z. Yoshida, J. Phys. A: Math. Theor. \textbf{48}, 205501 (2015).
\bibitem{bib07} N. Sato, Z. Yoshida, and Y. Kawazura, Plasma Fus. Res. \textbf{11}, 2401009 (2016).
\bibitem{bib08} T. J. Birmingham, T. G. Northrop, and C. G. F\"{a}lthammar, Phys. Fluids \textbf{10}, 11 (1967).
\bibitem{bib09} Z. Yoshida and S. M. Mahajan, Prog. Theor. Exp. Phys. \textbf{2014}, 073J01 (2014).
\bibitem{bib10} Z. Yoshida, Adv. Phys. X \textbf{1}, pp. 2-19 (2016).
\bibitem{bib11} N. Sato and Z. Yoshida, Phys. Rev. E \textbf{93}, 062140 (2016).
\bibitem{bib12} N. Sato, N. Kasaoka, and Z. Yoshida, Phys. Plasmas \textbf{22}, 042508 (2015). 
\bibitem{bib13} J. H. Malmberg and J. S. deGrassie, Phys. Rev. Lett. \textbf{35}, 9, pp. 577-580 (1975).
\bibitem{bib14} J. H. Malmberg and C. F. Driscoll, Phys. Rev. Lett. \textbf{44}, 10, pp. 654-657 (1980).
\bibitem{bib15} D. H. E. Dubin and T. M. O'Neil, Rev. Mod. Phys. \textbf{71}, 1, pp. 87-172 (1999).
\bibitem{bib16} S. A. Prasad and T. M. O'Neil, Phys. Fluids \textbf{22}, 278 (1979).
\bibitem{bib17} T. S. Pedersen and A. H. Boozer, Phys. Rev. Lett. \textbf{88}, 20 (2002).
\bibitem{bib18} H. Saitoh, J. Stanja, E. V. Stenson, U. Hergenhahn, H. Niemann, T. Sunn Pedersen, M. R. Stoneking, C. Piochacz, and C. Hugenschmidt, New J. Phys. \textbf{17}, 103038 (2015).
\bibitem{bib19} A. Hasegawa, L. Chen, and M. E. Mauel, Nucl. Fusion \textbf{30}, 11 (1990).
\bibitem{bib20} J. W. Gibbs, \textit{Elementary Principles in Statistical Mechanics} (Scribner's Sons, New York, 1902), pp. 3-19.
\bibitem{bib21} G. D. Birkhoff, Proc. Nat. Acad. Sci. U.S.A. \textbf{17}, 12, pp. 656-660 (1931).
\bibitem{bib22} Although non-canonical, equation (\ref{eq3}) is Hamiltonian because it can be written in the form $i_{\bol{v}}\mathcal{B}=-d\phi$, where $\mathcal{B}=d\mathcal{A}$ is the closed magnetic field $2$-form with potential 1-form $\mathcal{A}$, and $\phi$ serves as Hamiltonian function (remember that here $\bol{E}=-\nabla\phi$). 
\bibitem{bib23} P. J. Morrison, Rev. Mod. Phys. \textbf{70}, 2 (1998).
\bibitem{bib24} T. Frankel, \textit{The Geometry of Physics, An Introduction} (Cambridge University Press, Cambridge, 3rd ed., 2012), pp. 165-178.
\bibitem{bib25} C. W. Gardiner, \textit{Handbook of Stochastic Methods for Physics, Chemistry and the Natural Sciences} (Springer, 2nd ed., 1985).	
\bibitem{bib26}	For clarity of exposition in the derivation of the Fokker-Planck equation,
we have not discussed the mathematical issue associated with the definition of the stochastic integral. Here we adopted the Stratonovich integral, which is appropriate for an isolated system where random processes can be though as the limiting representation of a continuous perturbation. See \cite{bib06,bib25} for additional details.  
\end{thebibliography}

\end{document}